\documentclass{article}

\begin{document}

\title{\bf{Noncommutative sedeons and their application in field theory}}

\author{ Victor L. Mironov\thanks{E-mail: mironov@ipmras.ru}~ and Sergey V. Mironov   \\[15pt] Institute for physics of microstructures RAS, Nizhniy Novgorod, Russia }

\date{ (Submitted November 17, 2011, revised February 26, 2015)}

\maketitle

\begin{abstract}
   We present sixteen-component values "sedeons", generating
   associative noncommutative space-time algebra.
   The generalized second-order and first-order equations
   of relativistic quantum mechanics based on sedeonic wave
   function and sedeonic space-time operators are proposed.
   We also discuss the  description of fields with massive quantum
   on the basis of second-order and first-order equations for sedeonic potentials.
\end{abstract}



\section*{Introduction}

The application of multicomponent hypercomplex numbers and
multivectors in classical and quantum physics has a long history.
In particular, the simplest generalization of electrodynamics and
quantum mechanics was developed on the basis of
quaternions~\cite{b_1}-\cite{b_6}. The structure of quaternions
with four components (scalar and vector) corresponds to the
relativistic four-vector approach that allows one to reformulate
field equations in terms of quaternionic algebra. However, the
essential imperfection of the quaternionic algebra is that the
quaternions do not include pseudoscalar and pseudovector
components. The consideration of full space symmetry with respect
to spatial inversion leads to the eight-component structures
enclosing scalar, pseudoscalar, vector and pseudovector. There is
a lot of works on application of different eight-component values
such as biquaternions and octonions in classical electrodynamics
and relativistic quantum mechanics~\cite{b_7}-\cite{b_19}.
However, a consistent relativistic approach implies equally the
space and time symmetries that requires the consideration of the
extended sixteen-component space-time algebras.

There are a few approaches in the development of theory on the
basis of sixteen-component structures. One of them is the
application of hypercomplex numbers sedenions, which are obtained
from octonions by Cayley-Dickson extension procedure
\cite{b_20}-\cite{b_24}. But as in the case of octonions the
essential imperfection of sedenions is their nonassociativity.
Another approach is based on the application of hypercomplex
multivectors generating associative space-time Clifford algebras.
The basic idea of such multivectors is an introduction of
additional noncommutative time unit vector, which is orthogonal to
the space unit vectors \cite{b_25,b_26}. However, the application
of such multivectors in quantum mechanics is considered in general
as one of abstract algebraic scheme enabling the reformulation of
Klein-Gordon and Dirac equations for the multicomponent wave
functions but does not touch the physical entity of these
equations.

Recently we have developed an alternative approach based on our
scalar-vector concept \cite{b_27}-\cite{b_29} realized in
sixteen-component sedeons. In present paper we demonstrate the
application of the sedeons to the reformulation  of relativistic
quantum mechanics and massive field equations.

\section{Sedeonic space-time algebra}
\label{sec:1} The sedeonic algebra encloses four groups of values,
which are differed with respect to spatial and time inversion.

\begin{enumerate}
\item  Absolute scalars ($V$) and absolute vectors ($\vec{V}$) are
not transformed under spatial and time inversion.

\item  Time scalars ($V_{\bf t}$) and time vectors ($\vec{V}_{\bf
t}$) are changed (in sign) under time inversion but are not
transformed under spatial inversion.

\item  Space scalars ($V_{\bf r} $) and space vectors
($\vec{V}_{\bf r} $) are changed under spatial inversion but are
not transformed under time inversion.

\item  Space-time scalars ($V_{\bf tr} $) and space-time vectors
($\vec{V}_{\bf tr} $) are changed under spatial and time
inversion.
\end{enumerate}
Here indexes \textbf{t} and \textbf{r} indicate the
transformations (\textbf{t} for time inversion and \textbf{r} for
spatial inversion), which change the corresponding values. All
introduced values can be integrated into one space-time sedeon
$\tilde{{\bf V}}$, which is defined by the following expression:
\begin{equation}\label{eq_2_1}
\tilde{{\bf V}}=V+\vec{V}+V_{\bf t}+\vec{V}_{\bf t} +V_{\bf
r}+\vec{V}_{\bf r}+V_{\bf tr}+\vec{V}_{\bf tr}.
\end{equation}
Let us introduce scalar-vector basis $\bf a_0$, $\bf a_1$, $\bf
a_2$, $\bf a_3$, where the value ${\bf a_0}\equiv1$ is absolute
scalar unit and the values $\bf a_1$, $\bf a_2$, $\bf a_3$ are
absolute unit vectors generating the right Cartesian basis. We
introduce also four space-time scalar units  $\bf e_0$, $\bf e_1$,
$\bf e_2$, $\bf e_3$, where value ${\bf e_0}\equiv1$  is a
absolute scalar unit; ${\bf e_1}\equiv{\bf e_t}$ is a time scalar
unit; ${\bf e_2}\equiv{\bf e_r}$ is a space scalar unit; ${\bf
e_3}\equiv{\bf e_{tr}}$ is a space-time scalar unit. Using
space-time scalar units ${\bf e_j}$  (${\bf j} = 0, 1, 2, 3$) and
scalar-vector basis ${\bf a_k}$  (${\bf {k}}=0, 1, 2, 3$) we can
introduce unified sedeonic components $V_\mathrm{{jk}}$ in
accordance with the following relations:
\begin{equation}\label{eq_2_2}
\begin{array}{l}{V={\bf e_0}V_{00}{\bf a_0},}\\{\vec{V}={\bf e_0}\left(V_{01}{\bf a_1}+V_{02}{\bf a_2}+V_{03}{\bf a_3}\right),}\\{V_{\bf t}={\bf e_1}V_{10}{\bf a_0},}\\{\vec{V}_{\bf t}={\bf e_1}\left(V_{11}{\bf a_1}+V_{12}{\bf a_2}+V_{13}{\bf a_3}\right),}\\{V_{\bf r}={\bf e_2}V_{20}{\bf a_0},}\\{\vec{V}_{\bf r}={\bf e_2}\left(V_{21}{\bf a_1}+V_{22}{\bf a_2}+V_{23}{\bf a_3}\right),}\\{V_{\bf tr}={\bf e_3}V_{30}{\bf a_0},}\\{\vec{V}_{\bf tr}={\bf e_3}\left(V_{31}{\bf a_1}+V_{32}{\bf a_2}+V_{33}{\bf a_3}\right).}\end{array}
\end{equation}
Then the sedeon (\ref{eq_2_1}) can be written in the following
expanded form:
\begin{equation}\label{eq_2_3}
\begin{array}{c}{\tilde{{\bf V}}={\bf e_0}\left(V_{00}{\bf a_0}+V_{01}{\bf a_1}+V_{02}{\bf a_2}+V_{03}{\bf a_3}\right)}\\~~~~~{+{\bf e_1}\left(V_{10}{\bf a_0}+V_{11}{\bf a_1}+V_{12}{\bf a_2}+V_{13}{\bf a_3}\right)}\\~~~~~{+{\bf e_2}\left(V_{20}{\bf a_0}+V_{21}{\bf a_1}+V_{22}{\bf a_2}+V_{23}{\bf a_3}\right)}\\~~~~~~{+{\bf e_3}\left(V_{30}{\bf a_0}+V_{31}{\bf a_1}+V_{32}{\bf a_2}+V_{33}{\bf a_3}\right).}\end{array}
\end{equation}
The sedeonic components $V_\mathrm{jk}$  are numbers (complex in
general). Further we will use symbol $1$ instead of units $\bf
a_0$ and $\bf e_0$  for simplicity.

The multiplication and commutation rules for sedeonic absolute
unit vectors $\bf a_1$, $\bf a_2$, $\bf a_3$ and space-time units
$\bf e_1$, $\bf e_2$, $\bf e_3$ are presented in tables 1 and 2
respectively.

\begin{table}
\caption{} \label{tab:1}
\begin{center}
\begin{tabular}{llll}
\hline\noalign{\smallskip}
 & ~~$\bf a_1$ & ~~$\bf a_2$ & ~~$\bf a_3$ \\
\noalign{\smallskip}\hline\noalign{\smallskip}
$\bf a_1$ & ~~~1 & ~~$i\bf a_3$ & $-i\bf a_2$\\
$\bf a_2$ & $-i\bf a_3$ & ~~~1 & ~~$i\bf a_1$\\
$\bf a_3$ & ~~$i\bf a_2$ & $-i\bf a_1$& ~~~1 \\
\noalign{\smallskip}\hline
\end{tabular}
\end{center}

\caption{} \label{tab:2}
\begin{center}
\begin{tabular}{llll}
\hline\noalign{\smallskip}
 & ~~$\bf e_1$ & ~~$\bf e_2$ & ~~$\bf e_3$ \\
\noalign{\smallskip}\hline\noalign{\smallskip}
$\bf e_1$ & ~~~1 & ~~$i\bf e_3$ & $-i\bf e_2$\\
$\bf e_2$ & $-i\bf e_3$ & ~~~1 & ~~$i\bf e_1$\\
$\bf e_3$ & ~~$i\bf e_2$ & $-i\bf e_1$& ~~~1 \\
\noalign{\smallskip}\hline
\end{tabular}
\end{center}
\end{table}

In the tables and further the value ${\it i}$ is the imaginary
unit ($i^2=-1$). Note that sedeonic units  $\bf e_1$, $\bf e_2$,
$\bf e_3$  and unit vectors  $\bf a_1$, $\bf a_2$, $\bf a_3$
generate the anticommutative algebras:
\begin{equation}\label{eq_2_41}
\begin{array}{l}{\bf a_n\bf a_m=-\bf a_m\bf a_n},\\{\bf e_n\bf e_m=-\bf e_m\bf
e_n},
\end{array}
\end{equation}
for ${\bf {n}}$ and ${\bf m} =  1, 2, 3$ (${\bf n}\neq{\bf m}$),
but $\bf e_1$, $\bf e_2$, $\bf e_3$  commute with ${\bf a_1}$,
$\bf a_2$, ${\bf a_3}$:
\begin{equation}\label{eq_2_42}
\bf a_n\bf e_m=\bf e_m\bf a_n ,
\end{equation}
for any ${\bf {n}}$ and ${\bf m}$.

Thus the sedeon $\tilde{{\bf V}}$ is the complicated space-time
object consisting of absolute scalar, time scalar, space scalar,
space-time scalar, absolute vector, time vector, space vector and
space-time vector.

Introducing the designations of scalar-vector values
\begin{equation}\label{eq_2_4}
\begin{array}{l}{\overline{{\bf V}}_0=V_{00}+V_{01}{\bf a_1}+V_{02}{\bf a_2}+V_{03}{\bf a_3},}
\\{\overline{{\bf V}}_1=V_{10}+V_{11}{\bf a_1}+V_{12}{\bf a_2}+V_{13}{\bf a_3},}
\\{\overline{{\bf V}}_2=V_{20}+V_{21}{\bf a_1}+V_{22}{\bf a_2}+V_{23}{\bf a_3},}
\\{\overline{{\bf V}}_3=V_{30}+V_{31}{\bf a_1}+V_{32}{\bf a_2}+V_{33}{\bf a_3},}\end{array}
\end{equation}
we can write the sedeon (\ref{eq_2_3}) in the compact form
\begin{equation}\label{eq_2_5}
\tilde{{\bf V}}=\overline{{\bf V}}_0+{\bf e_1}\overline{{\bf
V}}_1+{\bf e_2}\overline{{\bf V}}_2+{\bf e_3}\overline{{\bf V}}_3.
\end{equation}
On the other hand, introducing the designations of space-time
sedeon-scalars
\begin{equation}\label{eq_2_6}
\begin{array}{l}{{\bf V}_0=V_{00}+{\bf e_1}V_{10}+{\bf e_2}V_{20}+{\bf e_3}V_{30},}
\\{{\bf V}_1=V_{01}+{\bf e_1}V_{11}+{\bf e_2}V_{21}+{\bf e_3}V_{31},}
\\{{\bf V}_2=V_{02}+{\bf e_1}V_{12}+{\bf e_2}V_{22}+{\bf e_3}V_{32},}
\\{{\bf V}_3=V_{03}+{\bf e_1}V_{13}+{\bf e_2}V_{23}+{\bf e_3}V_{33},}\end{array}
\end{equation}
we can write the sedeon (\ref{eq_2_3}) in another form
\begin{equation}\label{eq_2_7}
\tilde{{\bf V}}={\bf V}_0+{\bf V}_1{\bf a_1}+{\bf V}_2{\bf
a_2}+{\bf V}_3{\bf a_3},
\end{equation}
or introducing the sedeon-vector
\begin{equation}\label{eq_2_8}
\vec{{\bf V}}=\vec{V}+\vec{V}_{\bf t}+\vec{V}_{\bf r}
+\vec{V}_{\bf tr}={\bf V}_1{\bf a_1}+{\bf V}_2{\bf a_2}+{\bf
V}_3{\bf a_3},
\end{equation}
it can be represented in following compact form:
\begin{equation}\label{eq_2_9}
\tilde{{\bf V}}={\bf V}_0+\vec{{\bf V}}.
\end{equation}
Further we will indicate the sedeon-scalars and the sedeon-vectors
with the bold capital letters.

Let us consider the sedeonic multiplication in detail. The
sedeonic product of two sedeons $\tilde{{\bf A}}$ and $\tilde{{\bf
B}}$ can be presented in the following form:

\begin{equation}
\label{eq_2_10}
\begin{array}{l}{\tilde{{\bf A }} \tilde{{\bf B}} =\left({\bf
A}_{0} +\vec{{\bf A}}\right)\left({\bf B}_{0} +\vec{{\bf
B}}\right)}\\{={\bf A }_{0} {\bf B}_{0} +{\bf A}_{0} \vec{{\bf B
}} +\vec{{\bf A}}{\bf B}_{0} +\left(\vec{{\bf A}}\cdot \vec{{\bf
B}} \right)+\left[\vec{{\bf A}}\times \vec{{\bf B}}\right]}.
\end{array}
\end{equation}
Here we denote the sedeonic scalar multiplication of two
sedeon-vectors (internal product) by symbol ``$\cdot $'' and round
brackets
\begin{equation}
\label{eq_2_11} \left(\vec{{\bf A}}\cdot \vec{{\bf B}}
\right)={\bf A}_{1}{\bf B}_{1}+ {\bf A}_{2}{\bf B}_{2}+ {\bf
A}_{3}{\bf B}_{3},
\end{equation}
and sedeonic vector multiplication (external product) by symbol
``$\times$'' and square brackets

\begin{equation}
\label{eq_2_12} \begin{array}{l}\left[\vec{{\bf A}}\times
\vec{{\bf B}}\right]=i\left({\bf A }_{2} {\bf B}_{3} -{\bf A}_{3}
{\bf B}_{2}\right){\bf a_1} +i\left({\bf A}_{3} {\bf
B}_{1} -{\bf A}_{1} {\bf B}_{3}\right){\bf a_2} \\
~~~~~~~~~~~~+i\left({\bf A}_{1} {\bf B}_{2} -{\bf A}_{2} {\bf
B}_{1}\right){\bf a_3}.
\end{array}
\end{equation}

In (\ref{eq_2_11}) and (\ref{eq_2_12}) the multiplication of
sedeonic components is performed in accordance with (\ref{eq_2_6})
and table 2. Note that in sedeonic algebra the expression for the
vector product has some difference from analogous expression in
Gibbs vector algebra. Let us consider three absolute vectors
$\vec{A}$, $\vec{B}$  and $\vec{C}$. Then the formula for the
vector triple product in sedeonic algebra has the following form:
\begin{equation}
\label{eq_2_13} \left[\vec{{A}}\times \left[\vec{{B}}\times
\vec{{C}}\right]\right]=-\vec{{B}}
\left(\vec{{A}}\cdot\vec{{C}}\right)+
\vec{{C}}\left(\vec{{A}}\cdot\vec{{B}}\right).
\end{equation}

Thus, the sedeonic product
\begin{equation}\label{eq_2_14}
\tilde{{\bf F}}=\tilde{{\bf A}}\tilde{{\bf B}}={\bf F}_0+\vec{{\bf
F}}
\end{equation}has the following components:
\begin{equation}\label{eq_2_15}
\begin{array}{l}{{\bf F}_0={\bf A}_0{\bf B}_0+{\bf A}_1{\bf
B}_1+{\bf A}_2{\bf B}_2+{\bf A}_3{\bf B}_3,}\\{{\bf F}_1={\bf
A}_1{\bf B}_0+{\bf A}_0{\bf B}_1+i{\bf A}_2{\bf B}_3-i{\bf
A}_3{\bf B}_2,}\\{{\bf F}_2={\bf A}_2{\bf B}_0+{\bf A}_0{\bf
B}_2+i{\bf A}_3{\bf B}_1-i{\bf A}_1{\bf B}_3,}\\{{\bf F}_3={\bf
A}_3{\bf B}_0+{\bf A}_0{\bf B}_3+i{\bf A}_1{\bf B}_2-i{\bf
A}_2{\bf B}_1}.
\end{array}
\end{equation}

\section{Sedeonic spatial rotation and space-time conjugation}

The rotation of the sedeon $\tilde{{\bf V}}$ on the angle $\theta$
around the absolute unit vector $\vec{n}$ is realized by
uncompleted sedeon
\begin{equation}
\label{eq_3_1} \tilde{{\bf
U}}=\cos(\theta/2)+i\vec{n}\sin(\theta/2)
\end{equation}
and by complex conjugated sedeon
\begin{equation}
\label{eq_3_2} \tilde{{\bf
U}}^{*}=\cos(\theta/2)-i\vec{n}\sin(\theta/2),
\end{equation}
which satisfy the relation
\begin{equation}
\label{eq_3_3} \tilde{{\bf U}}^{*}\tilde{{\bf U}}=\tilde{{\bf
U}}\tilde{{\bf U}}^{*}=1.
\end{equation}
The transformed sedeon $\tilde{{\bf V}}^{'}$ is defined as the
sedeonic product
\begin{equation}
\label{eq_3_4} \tilde{{\bf V}}^{'}=\tilde{{\bf U}}^{*}\tilde{{\bf
V}}\tilde{{\bf U}}.
\end{equation}
Thus the transformed sedeon $\tilde{{\bf V}}^{'}$ can be written
in the following expanded form:
\begin{equation}
\label{eq_3_5}
\begin{array}{l}{\tilde{{\bf
V}}^{'}=\left (\cos(\theta/2)-i\vec{n}\sin(\theta/2)\right )({\bf
V}_0+\vec{\bf V})(\cos(\theta/2)+i\vec{n}\sin(\theta/2))}
\\[2mm]~~~~~{={\bf V}_0+\vec{\bf
V}\cos{\theta}+(1-\cos{\theta})(\vec{n}\cdot\vec{\bf
V})\vec{n}-i\sin{\theta}[\vec{n}\times\vec{\bf V}].}
\end{array}
\end{equation}
It is clearly seen that rotation does not transform the
sedeon-scalar part, but sedeonic vector $\vec{\bf V}$ is rotated
on the angle $\theta$ around $\vec{n}$.

The operations of time conjugation $(\widehat{R}_{\bf t})$, space
conjugation $(\widehat{R}_{\bf r})$ and space-time conjugation
$(\widehat{R}_{\bf tr})$ are connected with transformations in
${\bf e}_1$, ${\bf e}_2$, ${\bf e}_3$ basis and can be presented
as
\begin{equation}
\label{eq_3_6}
\begin{array}{l}{\widehat{R}_{\bf t}\tilde{\bf V}={\bf e_2}\tilde{\bf V}{\bf e_2}=\overline{\bf V}_0-{\bf e_1}\overline{\bf V}_1+{\bf e_2}\overline{\bf V}_2-{\bf e_3}\overline{\bf
V}_3},
\\{\widehat{R}_{\bf r}\tilde{\bf V}={\bf e_1}\tilde{\bf V}{\bf e_1}=\overline{\bf V}_0+{\bf e_1}\overline{\bf V}_1-{\bf e_2}\overline{\bf V}_2-{\bf e_3}\overline{\bf
V}_3},
\\{\widehat{R}_{\bf tr}\tilde{\bf V}={\bf e_3}\tilde{\bf V}{\bf e_3}=\overline{\bf V}_0-{\bf e_1}\overline{\bf V}_1-{\bf e_2}\overline{\bf V}_2+{\bf e_3}\overline{\bf
V}_3}.
\end{array}
\end{equation}

\section{Sedeonic Lorentz transformations}

The relativistic event four-vector can be represented in the
follow sedeonic form:
\begin{equation}
\label{eq_4_1} \tilde{{\bf S}}=i{\bf e_1}ct+{\bf e_2}\vec{r},
\end{equation} where $c$ is the velocity of light, $t$ is the absolute scalar of time and $\vec{r}$ is the absolute radius-vector.
The square of this value is the Lorentz invariant
\begin{equation}
\label{eq_4_2} \tilde{{\bf S}}\tilde{{\bf S}}=-c^2t^2+x^2+y^2+z^2.
\end{equation}

The Lorentz transformation of event four-vector is realized by
sedeons
\begin{equation}
\label{eq_4_3} \tilde{{\bf L}}=\cosh{\vartheta}-{\bf
e_3}\vec{m}\sinh{\vartheta},
\end{equation}
\begin{equation}
\label{eq_4_4} \tilde{\bf L}^{*}=\cosh{\vartheta}+{\bf
e_3}\vec{m}\sinh{\vartheta},
\end{equation}
where $\tanh{2\vartheta}=v/c$; $v$ is velocity of motion along the
absolute unit vector $\vec{m}$. Note that
\begin{equation}
\label{eq_4_5} \tilde{{\bf L}}^{*}\tilde{{\bf L}}=\tilde{{\bf
L}}\tilde{{\bf L}}^{*}=1.
\end{equation}
The transformed event four-vector $\tilde{{\bf S}}^{'}$ is written
as
\begin{equation}\label{eq_4_6}
\begin{array}{c} {\tilde{{\bf S}}^{'}=\tilde{{\bf L}}^{*}\tilde{{\bf
S}}\tilde{{\bf L}}=(\cosh\vartheta+{\bf
e_3}\vec{m}\sinh\vartheta)(i{\bf e_1}ct+{\bf
e_2}\vec{r})(\cosh\vartheta-{\bf
e_3}\vec{m}\sinh\vartheta)}\\[1mm]{=i{\bf e_1}ct\cosh2\vartheta-i{\bf
e_1}(\vec{m}\cdot\vec{r})\sinh2\vartheta}\\[1mm]{+{\bf
e_2}\vec{r}\cosh^2{\vartheta}-{\bf e_2}ct\vec{m}\sinh2\vartheta
}\\[1mm]{+{\bf e_2}(\vec{m}\cdot\vec{r})\vec{m}\sinh^2{\vartheta}+{\bf
e_2}[[\vec{m}\times\vec{r}]\times\vec{m}]\sinh^2\vartheta}.
\end{array}
\end{equation}
Separating the values with ${\bf e_1}$ and ${\bf e_2}$ we get the
well-known expressions for the time and coordinates
transformations \cite{b_30} :
\begin{equation}
\label{eq_4_7} t'=\frac{t-xv/c^2 }{\sqrt{1-v^2/c^2}},
~~~x'=\frac{x-tv}{\sqrt{1-v^2/c^2}},~~~y'=y,~~~z'=z,
\end{equation}
where $x$ is the coordinate along the $\vec{m}$ vector.

Let us also consider the Lorentz transformation of the full sedeon
$\tilde{\bf V}$. The transformed sedeon $\tilde{\bf V}^{'}$ can be
written as sedeonic product
\begin{equation}\label{eq_4_8} \tilde{{\bf
V}}^{'}=\tilde{{\bf L}}^{*}\tilde{{\bf V}}\tilde{{\bf L}}.
\end{equation}
In expanded form:
\begin{equation}\label{eq_4_9}
\begin{array}{c}{\tilde{{\bf V}}^{'}=(\cosh\vartheta+{\bf
e_{tr}}\vec{m}\sinh\vartheta)({\bf V}_0+\vec{\bf
V})(\cosh\vartheta-{\bf e_{tr}}\vec{m}\sinh\vartheta)}\\[1mm]{={\bf
V_0}\cosh^2\vartheta-{\bf e_{tr}}{\bf V}_0{\bf
e_{tr}}\sinh^2\vartheta+({\bf e_{tr}}{\bf V}_0-{\bf V}_0{\bf
e_{tr}})\vec{m}}\cosh\vartheta\sinh\vartheta\\[1mm]{+\vec{\bf
V}\cosh^2\vartheta-{\bf e_{tr}}\vec{m}\vec{\bf V}\vec{m}{\bf
e_{tr}}\sinh^2\vartheta+({\bf e_{tr}}\vec{m}\vec{\bf V}-\vec{\bf
V}\vec{m}{\bf e_{tr}})\cosh\vartheta\sinh\vartheta}.
\end{array}
\end{equation}
Rewriting the expression (\ref{eq_4_9}) with scalar
(\ref{eq_2_11}) and vector (\ref{eq_2_12}) products we get
\begin{equation}
\begin{array}{c}{\tilde{{\bf V}}^{'}={\bf
V_0}\cosh^2\vartheta-{\bf e_{tr}}{\bf V}_0{\bf
e_{tr}}\sinh^2\vartheta+({\bf e_{tr}}{\bf V}_0-{\bf V}_0{\bf
e_{tr}})\vec{m}\cosh\vartheta\sinh\vartheta}\\[1mm]{+\vec{\bf
V}\cosh^2\vartheta+{\bf e_{tr}}\vec{\bf V}{\bf
e_{tr}}\sinh^2\vartheta-2{\bf e_{tr}}(\vec{m}\cdot\vec{\bf V}){\bf
e_{tr}}\vec{m}\sinh^2\vartheta}\\[1mm]{+({\bf
e_{tr}}(\vec{m}\cdot\vec{\bf V})-(\vec{\bf V}\cdot\vec{m}){\bf
e_{tr}})\cosh\vartheta\sinh\vartheta}\\[1mm]{+({\bf
e_{tr}}[\vec{m}\times\vec{\bf V}]-[\vec{\bf V}\times\vec{m}]{\bf
e_{tr}})\cosh\vartheta\sinh\vartheta}.
\end{array}
\end{equation}
Thus, the transformed sedeon have the following components:
\begin{equation}
\begin{array}{l}{V^{'}=V},\\{V^{'}_{\bf tr}}=V_{\bf tr},
\\{{V^{'}_{\bf r}}=V_{\bf r}\cosh2\vartheta+{\bf e_{tr}}(\vec{m}\cdot\vec{V_{\bf t}})\sinh2\vartheta,}
\\{{V^{'}_{\bf t}}=V_{\bf t}\cosh2\vartheta+{\bf e_{tr}}(\vec{m}\cdot\vec{V_{\bf r}})\sinh2\vartheta,}
\\{\vec{V}^{'}=\vec {V}\cosh2\vartheta-2(\vec{m}\cdot\vec{V})\vec{m}\sinh^2\vartheta+{\bf e_{tr}}[\vec{m}\times\vec{V}_{\bf {tr}}]\sinh2\vartheta},
\\{\vec{V}^{'}_{\bf {tr}}=\vec {V}_{\bf {tr}}\cosh2\vartheta-2(\vec{m}\cdot\vec{V}_{\bf {tr}})\vec{m}\sinh^2\vartheta+{\bf
e_{tr}}[\vec{m}\times\vec{V}]\sinh2\vartheta},
\\{\vec{V}^{'}_{\bf {r}}=\vec {V}_{\bf {r}}+2(\vec{m}\cdot\vec{V}_{\bf r})\vec{m}\sinh^2\vartheta+{\bf e_{tr}}V_{\bf
t}\vec{m}\sinh2\vartheta},
\\{\vec{V}^{'}_{\bf {t}}=\vec {V}_{\bf {t}}+2(\vec{m}\cdot\vec{V}_{\bf t})\vec{m}\sinh^2\vartheta+{\bf e_{tr}}V_{\bf
r}\vec{m}\sinh2\vartheta}.
\end{array}
\end{equation}

\section{Sedeonic generalization of Klein-Gordon equation}

The wave function of relativistic particle satisfies an equation,
which is obtained from the Einstein relation between energy and
momentum
\begin{equation}
\label{eq_5 2} E^2-cp^2=m_0^{2} c^{4}
\end{equation}
by means of changing classical energy $E$ and momentum $\vec{p}$
 on corresponding quantum-mechanical operators:
\begin{equation}
\label{eq_5 3} \hat{E}=i\hbar \frac{\partial }{\partial t} ~~~{\rm
and}~~~ \hat{\vec{p}}=-i\hbar \vec{\nabla },
\end{equation}
 where $c$ is the speed of light, $m_0$ is the mass of particle, $\hbar$ is the Planck constant.
 The absolute vector of gradient has the following form:
\begin{equation}
\label{eq_5 4} \vec{\nabla }=\frac{\partial }{\partial x} {\bf
a}_{{\rm 1}}
 +\frac{\partial }{\partial y} {\bf a}_{{\rm 2}}
 +\frac{\partial }{\partial z} {\bf a}_{{\rm 3}} .
\end{equation}

In sedeonic algebra the Einstein relation (\ref{eq_5 2}) can be
written as
\begin{equation}
\label{eq_5 5} (i{\bf e_{t}}E+{\bf e_{r}}c\vec{p}+{\bf
e_{tr}}m_0c^2)(i{\bf e_{t}}E+{\bf e_{r}}c\vec{p}+{\bf
e_{tr}}m_0c^2)=0.
\end{equation}
Let us consider the wave function in the form of space-time sedeon
\begin{equation}
\label{eq_5 1} \tilde{{\bf V }}(\vec{r},t)={\bf V }_{0}
(\vec{r},t)+\vec{\bf V } (\vec{r},t).
\end{equation}
Then the generalized sedeonic wave equation for sedeonic wave
fucntion is written in the following form
\begin{equation}
\label{eq_5 6} \displaystyle \left(i{\bf e_{t}}\frac{1}{c}
\frac{\partial }{\partial t} -{\bf e_{r}}\vec{\nabla }-i{\bf
e_{tr}}\frac{m_0c}{\hbar}\right)\left(i{\bf e_{t}}\frac{1}{c}
\frac{\partial }{\partial t} -{\bf e_{r}}\vec{\nabla }-i{\bf
e_{tr}}\frac{m_0c}{\hbar}\right)\tilde{{\bf V }}= 0.
\end{equation}
In this equation  the basis elements ${\bf e_t}$, ${\bf e_r}$,
${\bf e_{tr}}$ and ${\bf a_1}$, ${\bf a_2}$, ${\bf a_3}$ play the
role of the space-time operators, which transform the sedeonic
wave function $\tilde{\bf V}$ by means of component permutation.
In fact the equation (\ref{eq_5 6}) is the system of 16 scalar
equations for each component of wave function.

Redefining the operators
\begin{equation}
\label{eq_5 7}
\begin{array}{l}
\displaystyle{\partial_{\bf t}={\bf e_{t}}\frac{1}{c}
\frac{\partial }{\partial t}},
\\[3mm]\displaystyle{\vec{\nabla}_{\bf
r}={\bf e_{r}}\vec{\nabla}}, \linebreak \\[1mm]\displaystyle{m_{\bf
tr}={\bf e_{tr}}\frac{m_0c}{\hbar}},
\end{array}
\end{equation}
we can rewrite the equation (\ref{eq_5 6}) in compact form:
\begin{equation}
\label{eq_5 8} \displaystyle \left(i\partial_{\bf t} - \vec{\nabla
}_{\bf r}-im_{\bf tr}\right)\left(i\partial_{\bf t} - \vec{\nabla
}_{\bf r}-im_{\bf tr}\right)\tilde{{\bf V }}= 0.
\end{equation}

Formally, the sedeonic equation (\ref{eq_5 8}) can be represented
in the form of the system of Maxwell-like first-order equations.
Let us consider the sequential action of operators. After the
action of the first operator in the left part of equation
(\ref{eq_5 8}) we obtain
\begin{equation}
\label{eq_5 11}
\begin{array}{l}{\left(i\partial_{\bf t} -
\vec{\nabla }_{\bf r}-im_{\bf tr}\right)\tilde{{\bf V }}=
i\partial_{\bf t}{\bf V}_0+i\partial_{\bf t}\vec{\bf
V}}\\[3mm]{-\vec{\nabla }_{\bf r}{\bf V}_0-\left(\vec{\nabla }_{\bf
r}\cdot\vec{\bf V}\right)-\left[\vec{\nabla }_{\bf
r}\times\vec{\bf V}\right]-im_{\bf tr}{\bf V}_0-im_{\bf
tr}\vec{\bf V}}.
\end{array}
\end{equation}
Introducing the scalar and vector values
\begin{equation}
\label{eq_5 12}{\bf E}_0 = i\partial_{\bf t}{\bf
V}_0-\left(\vec{\nabla }_{\bf r}\cdot\vec{\bf V}\right)-im_{\bf
tr}{\bf V}_0,
\end{equation}
\begin{equation}
\label{eq_5 13}\vec{\bf E} = i\partial_{\bf t}\vec{\bf
V}-\vec{\nabla }_{\bf r}{\bf V}_0-\left[\vec{\nabla }_{\bf
r}\times\vec{\bf V}\right]-im_{\bf tr}\vec{\bf V},
\end{equation}
the relation (\ref{eq_5 13}) is presented as
\begin{equation}
\label{eq_5 14}\left(i\partial_{\bf t} - \vec{\nabla }_{\bf
r}-im_{\bf tr}\right)\tilde{{\bf V }}={\bf E}_0 +\vec{\bf E}.
\end{equation}
Then the wave equation (\ref{eq_5 8}) can be rewritten in the
following form:
\begin{equation}
\label{eq_5 15'}\left(i\partial_{\bf t} - \vec{\nabla }_{\bf
r}-im_{\bf tr}\right)\left({\bf E}_0 +\vec{\bf E}\right)=0.
\end{equation}
Applying the operator $\left(i\partial_{\bf t} - \vec{\nabla
}_{\bf r}-im_{\bf tr}\right)$ to both parts of equation (\ref{eq_5
15'}) and separating sedeon-scalar and sedeon-vector parts we get
the wave equations for the values $\bf{E_0}$ and $\vec{\bf E}$:
\begin{equation}
\label{eq_5 151} \displaystyle \left(i\partial_{\bf t} -
\vec{\nabla }_{\bf r}-im_{\bf tr}\right)\left(i\partial_{\bf t} -
\vec{\nabla }_{\bf r}-im_{\bf tr}\right){{\bf E }_0}=0,
\end{equation}
\begin{equation}
\label{eq_5 16} \displaystyle \left(i\partial_{\bf t} -
\vec{\nabla }_{\bf r}-im_{\bf tr}\right)\left(i\partial_{\bf t} -
\vec{\nabla }_{\bf r}-im_{\bf tr}\right)\vec{\bf E } =0.
\end{equation}
On the other hand, performing sedeonic multiplication in
expression (\ref{eq_5 15'}) and separating sedeon-scalar and
sedeon-vector parts we obtain the Maxwell-like system of
first-order equations:
\begin{equation}
\label{eq_5 17}
\begin{array}{l}
i\partial_{\bf t}{\bf E}_0-\left(\vec{\nabla }_{\bf
r}\cdot\vec{\bf E}\right)-im_{\bf tr}{\bf E}_0=0, \\[3mm]
i\partial_{\bf t}\vec{\bf E}-\left[\vec{\nabla }_{\bf
r}\times\vec{\bf E}\right]-im_{\bf tr}\vec{\bf E}-\vec{\nabla
}_{\bf r}{\bf E}_0=0.
\end{array}
\end{equation}
As one can see the values  $\bf{E_0}$ and $\vec{\bf E}$ can be
interpreted as the quantum field intensities. These fields are
defined on the whole space and carry information about the
kinematic properties of the particle.

The equation (\ref{eq_5 6}) can be generalized for a particle in
an external electromagnetic field. Let us consider the charged
particle with electrical charge $e$. In this case we have to
change operators in (\ref{eq_5 6}) by
\begin{equation}
\frac{\partial }{\partial t} \to \frac{\partial }{\partial t}
+i\frac{e}{\hbar } {\it \varphi} , ~~\vec{\nabla }\to \vec{\nabla
}-i\frac{e}{\hbar c} \vec{A},
\end{equation}
where $\varphi$ is scalar potential and $\vec A$ is vector
potential of electromagnetic field. Then we obtain the following
wave equation
\begin{equation}
\label{eq_5 19}
\begin{array}{l}
\displaystyle \left(i{\bf e_{t}}\frac{1}{c} \frac{\partial
}{\partial t}-{\bf e_{t}}\frac{e}{\hbar c} {\it \varphi} -{\bf
e_{r}}\vec{\nabla} +i{\bf e_{r}}\frac{e}{\hbar c} \vec{A}-i{\bf
e_{tr}}\frac{m_0c}{\hbar}\right) \\[12pt] \displaystyle \otimes\left(i{\bf e_{t}}\frac{1}{c}
\frac{\partial }{\partial t}-{\bf e_{t}}\frac{e}{\hbar c} {\it
\varphi} -{\bf e_{r}}\vec{\nabla} +i{\bf e_{r}}\frac{e}{\hbar c}
\vec{A}-i{\bf e_{tr}}\frac{m_0c}{\hbar}\right)\tilde{{\bf V }}= 0.
\end{array}
\end{equation}
This equation describes the charged particle with spin 1/2 in an
external electromagnetic field \cite{b_31}.

\section{Sedeonic generalization of Dirac equation}

The sedeonic algebra enables the reformulation of the first-order
Dirac equation ~\cite{b_32} as a wave equation for the sedeonic
wave function:
\begin{equation}
\label{eq_6 1}\left(i{\bf e_{t}}\frac{1}{c} \frac{\partial
}{\partial t} -{\bf e_{r}}\vec{\nabla }-i{\bf
e_{tr}}\frac{mc}{\hbar}\right)\tilde{{\bf V }}= 0.
\end{equation}
In this equation  the basis elements ${\bf e_t}$, ${\bf e_r}$,
${\bf e_{tr}}$ and ${\bf a_1}$, ${\bf a_2}$, ${\bf a_3}$ play the
role of the space-time operators, which transform the sedeonic
wave function $\tilde{\bf V}$ by means of component permutation.
In fact, equation (\ref{eq_6 1}) describes the special quantum
field with zero field intensities ${\bf E_0}$ and ${\vec{\bf E}}$
(see expression (\ref{eq_5 14})).

The equations (\ref{eq_6 1}) can be generalized for a particle in
an external electromagnetic field. In this case we have
\begin{equation}
\label{eq_6 2} \displaystyle \left(i{\bf e_{t}}\frac{1}{c}
\frac{\partial }{\partial t}-{\bf e_{t}}\frac{e}{\hbar c} {\it
\Phi} -{\bf e_{r}}\vec{\nabla} +i{\bf e_{r}}\frac{e}{\hbar c}
\vec{A}-i{\bf e_{tr}}\frac{m_0c}{\hbar}\right)\tilde{{\bf V }}= 0.
\end{equation}
This equation describes the particle with spin 1/2 in an external
electromagnetic field \cite{b_31}.

\section{Sedeonic second-order equation for massive field}

\subsection{Homogeneous equation}

The Einstein relation between energy and momentum (\ref{eq_5 2})
allows another field interpretation. In this case  $E$, $\vec{p}$
and $m_0$ can be interpreted as energy, momentum and mass of a
quantum of field. Then the equation
\begin{equation}
\label{eq_5 61} \displaystyle \left(i{\bf e_{1}}\frac{1}{c}
\frac{\partial }{\partial t} -{\bf e_{2}}\vec{\nabla }-i{\bf
e_{3}}\frac{m_0c}{\hbar}\right)\left(i{\bf e_{1}}\frac{1}{c}
\frac{\partial }{\partial t} -{\bf e_{2}}\vec{\nabla }-i{\bf
e_{3}}\frac{m_0c}{\hbar}\right)\tilde{{\bf W }}= 0
\end{equation}
is the wave equation for the field potential $\tilde{{\bf W }}$
and relation
\begin{equation}
\label{eq_5 2a} E^2-c^2p^2=m_0^{2} c^{4}
\end{equation}
can be considered as the dispersion relation for the free wave of
massive field.

Let us introduce new operators
\begin{equation}
\label{eq_5 7}
\begin{array}{l}
\displaystyle{\partial=\frac{1}{c} \frac{\partial }{\partial t}},
\\[5mm]\displaystyle m=\frac{m_0c}{\hbar}.
\end{array}
\end{equation}
Then we can rewrite the equation (\ref{eq_5 61}) in compact form:
\begin{equation}
\label{eq_5 8a}  \left(i{\bf e_{1}}\partial - {\bf
e_{2}}\vec{\nabla }-i{\bf e_{3}}m\right)\left(i{\bf e_{1}}\partial
- {\bf e_{2}}\vec{\nabla }-i{\bf e_{3}}m\right)\tilde{{\bf W }}=
0.
\end{equation}
Let us choose the potential in the following form:
\begin{equation}
\label{eq_5 9} \tilde{{\bf W }}= a+i{\bf e_{1}}b-i{\bf
e_{2}}c-i{\bf e_{3}}d+i\vec A+{\bf e_{1}}\vec B+{\bf e_{2}}\vec
C-{\bf e_{3}}\vec D,
\end{equation}
where the components $a, b, c, d, \vec {A},\vec {B}, \vec {C}$ and
$\vec {D} $ are the functions of spatial coordinates and time.
Introducing the scalar and vector fields strengths according to
the following definitions:
\begin{equation}
\label{eq_5 10}
\begin{array}{l}
e=\partial b+(\vec\nabla\cdot\vec C)+md,
\\[1mm] f=\partial a+(\vec\nabla\cdot\vec D)+mc,
\\[1mm] g=\partial d+(\vec\nabla\cdot\vec A)-mb,
\\[1mm] h=\partial c+(\vec\nabla\cdot\vec B)-ma,
\\[1mm] \vec E=-\partial\vec B - \vec\nabla c-i[\vec\nabla\times\vec C]-m\vec D,
\\[1mm] \vec F=-\partial\vec A - \vec\nabla d+i[\vec\nabla\times\vec D]-m\vec C,
\\[1mm] \vec G=-\partial\vec D - \vec\nabla a-i[\vec\nabla\times\vec A]+m\vec B,
\\[1mm] \vec H=-\partial\vec C - \vec\nabla b+i[\vec\nabla\times\vec B]+m\vec A,
\end{array}
\end{equation}
we get
\begin{equation}
\label{eq_5 11}
\begin{array}{l}
\left(i{\bf e_{1}}\partial - {\bf e_{2}}\vec{\nabla }-i{\bf
e_{3}}m\right)\left(a+i{\bf e_{1}}b-i{\bf e_{2}}c-i{\bf
e_{3}}d+i\vec A+{\bf e_{1}}\vec B+{\bf e_{2}}\vec C-{\bf
e_{3}}\vec D \right)
\\[5mm] =-e+i{\bf e_{1}}f-i{\bf
e_{2}}g+i{\bf e_{3}}h-i\vec E+{\bf e_{1}}\vec F+{\bf e_{2}}\vec
G+{\bf e_{3}}\vec H
\end{array}
\end{equation}
and the wave equation (\ref{eq_5 8a}) takes the form
\begin{equation}
\label{eq_5 12} \left(i{\bf e_{1}}\partial - {\bf
e_{2}}\vec{\nabla }-i{\bf e_{3}}m\right)\left(-e+i{\bf
e_{1}}f-i{\bf e_{2}}g+i{\bf e_{3}}h-i\vec E+{\bf e_{1}}\vec F+{\bf
e_{2}}\vec G+{\bf e_{3}}\vec H\right)=0.
\end{equation}
Performing the action of operator in the left part of the equation
(\ref{eq_5 12}), and separating the terms with different
space-time properties, we obtain the system of equations for the
field's strengths, similar to the system of Maxwell's equations in
electrodynamics:
\begin{equation}
\label{eq_5 13a}
\begin{array}{l}
        \partial f+(\vec\nabla\cdot\vec G)-mh=0,
\\[1mm] \partial e+(\vec\nabla\cdot\vec H)-mg=0,
\\[1mm] \partial h+(\vec\nabla\cdot\vec E)+mf=0,
\\[1mm] \partial g+(\vec\nabla\cdot\vec F)+me=0,
\\[1mm] \partial\vec F + \vec\nabla g+i[\vec\nabla\times\vec G]-m\vec H=0,
\\[1mm] \partial\vec E + \vec\nabla h-i[\vec\nabla\times\vec H]-m\vec G=0,
\\[1mm] \partial\vec H + \vec\nabla e+i[\vec\nabla\times\vec E]+m\vec F=0,
\\[1mm] \partial\vec G + \vec\nabla f-i[\vec\nabla\times\vec F]+m\vec E=0.
\end{array}
\end{equation}

The proposed equations for massive field possess a specific gauge
invariance. It is easy to see that fields strengths (\ref{eq_5
10}) and equations (\ref{eq_5 13a}) are not changed under the
following substitutions for potentials:
\begin{equation}
\label{gi}
\begin{array}{l}
a \Rightarrow a+\partial\varepsilon_a -m\varepsilon_c,
\\[1mm] b \Rightarrow b+\partial\varepsilon_b -m\varepsilon_d,
\\[1mm] c \Rightarrow c+\partial\varepsilon_c + m\varepsilon_a,
\\[1mm] d \Rightarrow d+\partial\varepsilon_d + m\varepsilon_b,
\\[1mm] \vec A \Rightarrow \vec A - \vec\nabla \varepsilon_d,
\\[1mm] \vec B \Rightarrow \vec B - \vec\nabla \varepsilon_c,
\\[1mm] \vec C \Rightarrow \vec C - \vec\nabla \varepsilon_b,
\\[1mm] \vec D \Rightarrow \vec D - \vec\nabla
\varepsilon_a,
\end{array}
\end{equation}
where $\varepsilon_a$, $\varepsilon_b$, $\varepsilon_c$,
$\varepsilon_d$, are arbitrary scalar functions, which satisfy
homogeneous Klein-Gordon equation. These gauge conditions are
different from those taken in electrodynamics \cite{b_33}.

 Multiplying each of
the equations (\ref{eq_5 13a}) to the corresponding field strength
and adding these equations to each other, we obtain:
\begin{equation}
\label{eq_5 14a}
\begin{array}{l}
\displaystyle{\frac{1}{2}\partial}\left(f^2+e^2+h^2+g^2+\vec
F^2+\vec E^2+\vec H^2+\vec G^2 \right)
\\[4mm] +f\left(\vec\nabla\cdot\vec G\right)
+e\left(\vec\nabla\cdot\vec H\right)+h\left(\vec\nabla\cdot\vec
E\right)+g\left(\vec\nabla\cdot\vec F\right) \\[4mm] +\left(\vec F\cdot\vec\nabla g\right)+\left(\vec E\cdot\vec\nabla h\right)
+\left(\vec H\cdot\vec\nabla e\right)+\left(\vec G\cdot\vec\nabla
f\right)
\\[4mm] +i \left(\vec F\cdot[\vec\nabla\times\vec G]\right)-i \left(\vec E\cdot[\vec\nabla\times\vec H]\right)
\\[4mm] +i \left(\vec H\cdot[\vec\nabla\times\vec E]\right)-i \left(\vec
G\cdot[\vec\nabla\times\vec F]\right)=0.
\end{array}
\end{equation}
Let us introduce the following notations:
\begin{equation}
\label{eq_5 15}
w=\displaystyle{-\frac{1}{8\pi}}\left(f^2+e^2+h^2+g^2+\vec F^2+\vec
E^2+\vec H^2+\vec G^2 \right),
\end{equation}
\begin{equation}
\label{eq_5 16} \vec P=\displaystyle{-\frac{c}{4\pi}}\left(e\vec
H+f\vec G+g\vec F +h\vec E +i\left[\vec E\times\vec
H\right]+i\left[\vec G\times\vec F\right]\right).
\end{equation}
Then the equation (\ref{eq_5 14a}) can be written as:
\begin{equation}
\label{eq_5 17} \frac{1}{c} \frac{\partial w}{\partial
t}+\left(\vec\nabla\cdot\vec P\right)=0.
\end{equation}
This expression is an analog of the Poynting theorem for massive
field. The value $w$ plays the role of the field energy density
and $ \vec P$ is a vector of energy flux density. The minus sign
in expressions (\ref{eq_5 15}) and (\ref{eq_5 16}) are chosen with
respect to the attractive character of charge interaction (see
further Section 6.2.).

\subsection {Nonhomogeneous equation}

Let us consider the sedeonic nonhomogeneous equation for massive
field
\begin{equation}
\label{eq_5 18}  \left(i{\bf e_{1}}\partial - {\bf
e_{2}}\vec{\nabla }-i{\bf e_{3}}m\right)\left(i{\bf e_{1}}\partial
- {\bf e_{2}}\vec{\nabla }-i{\bf e_{3}}m\right)\tilde{{\bf W
}}=\tilde{{\bf J}},
\end{equation}
where $\tilde{{\bf J}}$ is the source of massive field. By analogy
with electrodynamics  we consider the source in the following form
\cite{b_28}
\begin{equation}
\label{eq_5 19}\tilde{\bf J}=-i{\bf e_{1}}4\pi\rho_B- {\bf
e_{2}}\frac{4\pi}{c}\vec j_B,
\end{equation}
where $\rho_B$ is a volume density of charge and $\vec j_B$ is
density of current. In this case we can describe the field by
sedeonic potential $\tilde{\bf W }$ written in the following form
\begin{equation}
\label{eq_5 19}\tilde{\bf W}=i{\bf e_{1}}b+ {\bf e_{2}}\vec C,
\end{equation}
where $b(\vec r, t)$  is a scalar part and $\vec C(\vec r, t)$ is
a vector part of field potential. In this case we have only the
following nonzero field's strengths
\begin{equation}
\label{eq_5 20}
\begin{array}{l}
e=\partial b+\left(\vec\nabla\cdot\vec C\right),
\\[2mm] g=-mb,
\\[2mm] \vec E=-i\left[\vec\nabla\times\vec C\right],
\\[2mm] \vec F=-m\vec C,
\\[2mm] \vec H=-\partial\vec C - \vec\nabla b,
\end{array}
\end{equation}
and the equation (\ref{eq_5 18}) can be rewritten as
\begin{equation}
\label{eq_5 201}
\begin{array}{l}\left(i{\bf e_{1}}\partial - {\bf
e_{2}}\vec{\nabla }-i{\bf e_{3}}m\right)\left(-e-i{\bf
e_{2}}g-i\vec E+{\bf e_{1}}\vec
F+{\bf e_{3}}\vec H\right) \\
\displaystyle=-i{\bf e_{1}}4\pi\rho_B- {\bf
e_{2}}\frac{4\pi}{c}\vec j_B.
\end{array}
\end{equation}
Then we obtain the following equations for the field strengths:
\begin{equation}
\label{eq_5 21}
\begin{array}{l}
        \partial e+(\vec\nabla\cdot\vec H)-mg=4\pi\rho_B,
\\[1mm] (\vec\nabla\cdot\vec E)=0,
\\[1mm] \partial g+(\vec\nabla\cdot\vec F)+me=0,
\\[1mm] \partial\vec F + \vec\nabla g-m\vec H=0,
\\[1mm] \partial\vec E -i[\vec\nabla\times\vec H]=0,
\\[1mm] \displaystyle\partial\vec H + \vec\nabla e+i[\vec\nabla\times\vec E]+m\vec F=-\frac {4\pi}{c}\vec j_B,
\\[1mm] i[\vec\nabla\times\vec F]-m\vec E=0.
\end{array}
\end{equation}
On the other hand, applying the operator $(i{\bf e_{1}}\partial -
{\bf e_{2}}\vec{\nabla }-i{\bf e_{3}}m)$ to the equation
(\ref{eq_5 201}) we obtain the following wave equations for the
field strengths:
\begin{equation}
\label{eq_5 211}
\begin{array}{l}
        \displaystyle(\partial^2-\triangle+m^2) e=4\pi(\partial\rho_B+\frac {1}{c}(\vec\nabla\cdot \vec j_B)),
\\[2mm] (\partial^2-\triangle+m^2)g=-4\pi m \rho_B,
\\[2mm] \displaystyle(\partial^2-\triangle+m^2)\vec F = -\frac {4\pi}{c}m\vec j_B,,
\\[2mm] \displaystyle(\partial^2-\triangle+m^2)\vec E =-i\frac {4\pi}{c}[\vec\nabla\times\vec j_B],
\\[2mm] \displaystyle(\partial^2-\triangle+m^2)\vec H =-4\pi(\frac {1}{c}\partial\vec j_B+\vec\nabla \rho_B).
\end{array}
\end{equation}
Assuming the charge conservation
\begin{equation}
\label{eq_5 212}\partial\rho_B+\frac {1}{c}(\vec\nabla\cdot \vec
j_B)=0
\end{equation}
we can choose the field strength $e$ equal to zero. This is
equivalent to the following gauge condition (see (\ref{eq_5 20})):
\begin{equation}
\label{eq_5 213} \partial b+(\vec\nabla\cdot\vec C)=0
\end{equation}
similar to the Lorentz gauge.

Let us consider the simplest case of stationary field of point
scalar source. In the stationary case $\vec j_B = 0$ and field
potential can be chosen in a scalar form
\begin{equation}
\label{eq_5 22}\tilde{\bf W}=i{\bf e_{1}}b\left(\vec{r}\right).
\end{equation}
Then we have only two nonzero field components
\begin{equation}
\label{eq_5 23}
\begin{array}{l}
g=-mb,
\\[2mm] \vec H= - \vec\nabla b
\end{array}
\end{equation}
and the following field equations:
\begin{equation}
\label{eq_5 24}
\begin{array}{l}
        (\vec\nabla\cdot\vec H)-mg=4\pi\rho_B,
\\[2mm] \vec\nabla g-m\vec H=0,
\\[2mm] [\vec\nabla\times\vec H]=0.
\end{array}
\end{equation}

Let us calculate the field produced by a scalar stationary point
 source
\begin{equation}
\label{eq_5 25}\tilde{\bf J}=-4\pi q_B\delta (\vec r),
\end{equation}
where $q_B$ is the point charge and $\delta (\vec r)$ is delta
function. Then stationary wave equation can be written in
spherical coordinates as
\begin{equation}
\label{eq_5 26} \left(\frac{1}{r^2} \frac{\partial}{\partial
r}\left(r^2\frac{\partial}{\partial r}\right)-
\frac{{m^2_0}c^2}{\hbar^2}\right)b(r)=-4\pi q_B\delta(\vec{r}).
\end{equation}
The partial solution of the equation (\ref{eq_5 26}), which decays
at $r\to\infty$, is
\begin{equation}
\label{eq_5 27} b=\frac{q_B}{r}\exp\left(-
\frac{{m_0}c}{\hbar}r\right).
\end{equation}
Thus in this case the stationary field has scalar and vector
components
\begin{equation}
\label{eq_5 28} g=\frac{{m_0}c}{\hbar}\frac{q_B}{r}\exp\left(-
\frac{{m_0}c}{\hbar}r\right),
\end{equation}
\begin{equation}
\label{eq_5 29} \vec
H=\left(\frac{1}{r}+\frac{{m_0}c}{\hbar}\right)\frac{q_B}{r}\exp\left(-
\frac{{m_0}c}{\hbar}r\right){\vec r}_0,
\end{equation}
where $\vec r_0$ is a unit radial vector.

Two point charges interact due to the overlap of their fields.
Taking into account that the field in this case is the sum of the
two fields $g=g_1+g_2$ and $\vec H=\vec H_1 + \vec H_2$ the energy
of interaction is equal (see (\ref{eq_5 15}))
\begin{equation}
\label{eq_5 30} W_{BB}=-\frac{1}{4\pi}\int \{g_1g_2+(H_1\cdot
H_2)\}dV,
\end{equation}
where the integral is over all space. This expression can be
derived analytically:
\begin{equation}
\label{eq_5 31} W_{BB}=-\frac{q_{B1}q_{B2}}{R}\exp\left(-
\frac{{m_0}c}{\hbar}R\right),
\end{equation}
where $R$ is the distance between the point charges. This
expression coinsides with a well-known law of interaction between
two baryons, which is described by Yukawa potential \cite{b_34},
therefore $q_B$ can be interpreted as a baryon charge.

\section {Sedeonic first-order equation for massive field}

\subsection {Homogeneous equation}
Let us consider a special massive field that is described by
sedeonic first-order equation. In sedeonic algebra the homogeneous
first-order Dirac-like equation corresponding to the equation
(\ref{eq_5 61}) is written as
\begin{equation}
\label{eq_6 1a}  \left(i{\bf e_{1}}\partial - {\bf
e_{2}}\vec{\nabla }-i{\bf e_{3}}m\right)\tilde{{\bf W }}= 0.
\end{equation}
Choosing potential in the form (\ref{eq_5 9}) we find that
sedeonic equation (\ref{eq_6 1a}) is equivalent to the following
system
\begin{equation}
\label{eq_6 2}
\begin{array}{l}
 \partial a+(\vec\nabla\cdot\vec D)+mc=0,
\\[1mm] \partial b+(\vec\nabla\cdot\vec C)+md=0,
\\[1mm] \partial c+(\vec\nabla\cdot\vec B)-ma=0,
\\[1mm] \partial d+(\vec\nabla\cdot\vec A)-mb=0,
\\[1mm] \partial\vec A + \vec\nabla d-i[\vec\nabla\times\vec D]+m\vec C=0,
\\[1mm] \partial\vec B + \vec\nabla c+i[\vec\nabla\times\vec C]+m\vec D=0,
\\[1mm] \partial\vec C + \vec\nabla b-i[\vec\nabla\times\vec B]-m\vec A=0,
\\[1mm] \partial\vec D + \vec\nabla a+i[\vec\nabla\times\vec A]-m\vec B=0.
\end{array}
\end{equation}
In fact, these equations describe the special field with zero
field strengths (see for comparison the expressions (\ref{eq_5
10})).

Let us consider the plane wave solution of equation (\ref{eq_6
1a}) in detail. In this case the potential can be written as
\begin{equation}\label{eq_7 1}\tilde{{\bf W }}=\tilde{{\bf U}}\; \exp\left\{-i{\omega}t+i\left(\vec{k}\cdot \vec{r}\right)\right\},
\end{equation}
where $\omega$ is a frequency and $\vec k$ is an absolute wave
vector; the amplitude of the wave $\bf U$ does not depend on the
coordinates and time. In this case, the dependence of frequency on
the wave vector has two branches:
\begin{equation}\label{eq 7 2}
\omega_\pm=\pm\sqrt{c^2k^2+\frac{m^2_0c^4}{\hbar^2}}~.
\end{equation}
Let us consider the amplitude of the wave function in the form of
(\ref{eq_5 9}):
\begin{equation}
\label{eq_7 3}\tilde{\bf U}=a+i{\bf e_{1}}b-i{\bf e_{2}}c-i{\bf e_{3}}d+i\vec{A}+ {\bf e_{1}}\vec B+{\bf e_{2}}\vec C-{\bf
e_{3}}\vec D,
\end{equation}
where $a$, $b$, $c$, $d$, $\vec A$, $\vec B$, $\vec C$ and $\vec D$ are arbitrary constants. Then the solution can be written as
\begin{equation}\label{eq_7 4}
\tilde{{\bf W }}=\left(a+i{\bf e_{1}}b-i{\bf e_{2}}c-i{\bf e_{3}}d+i\vec{A}+ {\bf e_{1}}\vec B+{\bf e_{2}}\vec C-{\bf
e_{3}}\vec D\right)
\exp\left\{-i{\omega}_\pm t+i\left(\vec{k}\cdot
\vec{r}\right)\right\}.
\end{equation}
Substituting this expression in the original equation (\ref{eq_6
1}) we get:
\begin{equation}\label{eq_7 5}
\left({\bf e_{1}}\frac{{\omega}_\pm}{c}-i {\bf e_{2}}\vec{k}-i{\bf
e_{3}}\frac {m_0 c}{\hbar}\right)\left(a+i{\bf e_{1}}b-i{\bf e_{2}}c-i{\bf e_{3}}d+i\vec{A}+ {\bf e_{1}}\vec B+{\bf e_{2}}\vec C-{\bf
e_{3}}\vec D\right)=0.
\end{equation}
For convenience we introduce the following notation:
\begin{equation}
\label{eq_7 6}
\begin{array}{l}\displaystyle\omega'=\frac{\omega_\pm}{c},
\\[3mm] \displaystyle m=\frac {m_0 c}{\hbar},
\end{array}
\end{equation}
then equation (\ref{eq_7 5}) can be rewritten as
\begin{equation}\label{eq_7 6}
\left({\bf e_{1}}\omega'-i {\bf e_{2}}\vec{k}-i{\bf
e_{3}}m\right)\left(a+i{\bf e_{1}}b-i{\bf e_{2}}c-i{\bf
e_{3}}d+i\vec{A}+ {\bf e_{1}}\vec B+{\bf e_{2}}\vec C-{\bf
e_{3}}\vec D\right)=0.
\end{equation}
 For fixed $\vec{k}$ let us represent
the vector constants in (\ref{eq_7 3}) in the form
\begin{equation}
\label{Const_PP}
\begin{array}{l}\displaystyle \vec A=\vec A_\parallel+\vec A_\perp, \\[3mm] \displaystyle \vec B=\vec B_\parallel+\vec B_\perp, \\[3mm]\displaystyle \vec C=\vec C_\parallel+\vec C_\perp, \\[3mm]\displaystyle \vec D=\vec D_\parallel+\vec D_\perp,
\end{array}
\end{equation}
where the vectors $\vec A_\parallel$, $\vec B_\parallel$, $\vec
C_\parallel$ and $\vec D_\parallel$ are parallel to the vector
$\vec k$ while the vectors $\vec A_\perp$, $\vec B_\perp$, $\vec
C_\perp$ and $\vec D_\perp$ are perpendicular to $\vec k$. Then
performing the multiplication in (\ref{eq_7 6}), we obtain the
following system of algebraic equations:
\begin{equation}
\label{s1}
i\omega'b-ikC_\parallel-md=0,
\end{equation}
\begin{equation}
\label{s2}
\omega'a-kD_\parallel+imc=0,
\end{equation}
\begin{equation}
\label{s3}
-\omega'd+kA_\parallel+imb=0,
\end{equation}
\begin{equation}
\label{s4}
\omega'c-kB_\parallel-ima=0,
\end{equation}
\begin{equation}
\label{s5}
\omega'B_\parallel-kc+imD_\parallel=0,
\end{equation}
\begin{equation}
\label{s6}
i\omega'A_\parallel-ikd-mC_\parallel=0,
\end{equation}
\begin{equation}
\label{s7}
i\omega'D_\parallel-ika+mB_\parallel=0,
\end{equation}
\begin{equation}
\label{s8}
i\omega'C_\parallel-ikb+mA_\parallel=0,
\end{equation}
\begin{equation}
\label{s9}
\omega'\vec B_\perp-i\left[\vec k\times \vec C_\perp \right]+im\vec D_\perp=0,
\end{equation}
\begin{equation}
\label{s10}
i\omega'\vec A_\perp-\left[\vec k\times \vec D_\perp \right]-m\vec C_\perp=0,
\end{equation}
\begin{equation}
\label{s11}
i\omega'\vec D_\perp+\left[\vec k\times \vec A_\perp \right]+m\vec B_\perp=0,
\end{equation}
\begin{equation}
\label{s12}
i\omega'\vec C_\perp-\left[\vec k\times \vec B_\perp \right]+m\vec A_\perp=0,
\end{equation}
where the values $A_\parallel$, $B_\parallel$, $C_\parallel$ and $D_\parallel$ are the projections of the vectors $\vec A_\parallel$, $\vec B_\parallel$, $\vec C_\parallel$ and $\vec D_\parallel$ on the vector $\vec k$.

Let us solve this system of equations. From (\ref{s11}) and
(\ref{s12}) we find
\begin{equation}
\label{sD} \vec D_\perp=\frac{im}{\omega'}\vec
B_\perp+\frac{i}{\omega'}\left[\vec k\times \vec A_\perp \right],
\end{equation}
\begin{equation}
\label{sA} \vec C_\perp=\frac{im}{\omega'}\vec
A_\perp-\frac{i}{\omega'}\left[\vec k\times \vec B_\perp \right].
\end{equation}
Using (\ref{eq 7 2}) one can easily check that for arbitrary
vector constants $A_\perp$ and $B_\perp$ equations (\ref{s9}) and
(\ref{s10}) are fulfilled.

As a next step from equations (\ref{s1})-(\ref{s4}) we obtain:
\begin{equation}
\label{f1}
C_\parallel=\frac{\omega'}{k}b+i\frac{m}{k}d,
\end{equation}
\begin{equation}
\label{f2}
D_\parallel=\frac{\omega'}{k}a+i\frac{m}{k}c,
\end{equation}
\begin{equation}
\label{f3}
A_\parallel=\frac{\omega'}{k}d-i\frac{m}{k}b,
\end{equation}
\begin{equation}
\label{f4}
B_\parallel=\frac{\omega'}{k}c-i\frac{m}{k}a.
\end{equation}
One can check that these solution fulfill the equations
(\ref{s5})-(\ref{s8}).

Thus the sedeon $\tilde{\bf U}$ has the form
\begin{equation}
\label{U_res}
\begin{array}{c}{\displaystyle
\tilde{\bf U}=a+i{\bf e_{1}}b-i{\bf e_{2}}c-i{\bf
e_{3}}d}\\{\displaystyle +\left\{i\omega'd+mb+ {\bf
e_{1}}\omega'c-i{\bf e_{1}}ma+{\bf e_{2}}\omega'b+i{\bf
e_{2}}md-{\bf e_{3}}\omega'a-i{\bf e_{3}}mc\right\}\frac{\vec
k}{k^2}}
\\[10pt]{\displaystyle +i\vec A_\perp+ {\bf e_{1}}\vec B_\perp+i{\bf
e_{2}}\frac{m}{\omega'}\vec A_\perp-i{\bf
e_{3}}\frac{m}{\omega'}\vec B_\perp}
\\[10pt] \displaystyle{-i{\bf
e_{3}}\frac{1}{\omega'}\left[\vec k\times \vec A_\perp
\right]-i{\bf e_{2}}\frac{1}{\omega'}\left[\vec k\times \vec
B_\perp \right].}
\end{array}
\end{equation}
Note that this expression can be rewritten in the following form:
\begin{equation}
\begin{array}{c}
\label{U_res_short} \tilde{\bf U}=\left({\bf e_{1}}\omega'-i {\bf
e_{2}}\vec k-i{\bf e_{3}}m\right)
\\ \displaystyle\times\left\{i{\bf e_{2}}\frac{\vec
k}{k^2}\left(a+i{\bf e_{1}}b-i{\bf e_{2}}c-i{\bf
e_{3}}d\right)+i{\bf e_{1}}\frac{1}{\omega'}\vec
A_\perp+\frac{1}{\omega'}\vec B_\perp\right\}.
\end{array}
\end{equation}
Substituted this amplitude into (\ref{eq_7 6}) one can see that
this equation is satisfied for any parameters $a,~b,~c,~d,~\vec
A_\perp,~\vec B_\perp $ because the expression in round brackets
is sedeonic zero divisor. Indeed it is simple to check that
\begin{equation}
\label{Zero_Div} \left({\bf e_{1}}\omega'-i {\bf e_{2}}\vec
k-i{\bf e_{3}}m\right)\left({\bf e_{1}}\omega'-i {\bf e_{2}}\vec
k-i{\bf e_{3}}m\right)\equiv 0.
\end{equation}
In general, the plane wave solution for the equation (\ref{eq_6
1a}) can be written in the following sedeonic form:
\begin{equation}
\label{U_res_gen} \tilde{\bf W}=\left({\bf e_{1}}\omega'-i {\bf
e_{2}}\vec k-i{\bf e_{3}}m\right)\tilde{\bf
M}\exp\left\{-i{\omega}t+i\left(\vec{k}\cdot
\vec{r}\right)\right\},
\end{equation}
where $\tilde{\bf M}$ is an arbitrary sedeon with constant
components. In this case after performing multiplication in
(\ref{U_res_gen}) we obtain that the components of the resulting
sedeon are defined only by 8 independent combinations of the
sedeon $\tilde{\bf M}$ components. Note that the internal
structure of this wave is changed under space and time inversion.

In massless case the dispersion relation is
\begin{equation}\label{eq 17 2}
\omega_\pm=\pm~{ck}
\end{equation}
and plane wave solution can be written as
\begin{equation}\label{eq 6 13}\tilde{{\bf W }}=\left({{\bf {e_t}}\frac{\omega_\pm}{c}}-i{\bf {e_r}}\vec{k}\right)\tilde{\bf
M}\exp\left\{-i{\omega_\pm}t+i\left(\vec{k}\cdot
\vec{r}\right)\right\}.
\end{equation}
Let us analyze the structure of the plane wave (\ref{eq 6 13}) in
detail. We suppose that wave vector is directed along z axis. Then
the first-order equation (\ref{eq_6 1a}) can be rewritten in the
following equivalent form:
\begin{equation} \label{eq_6_16_}
\left(\frac{1}{c} \frac{\partial }{\partial t} + {\bf e_{tr}}{\bf
a_3} \frac{\partial }{\partial z}\right)\tilde{{\bf W'}} =0,
\end{equation}
where $\tilde{\bf W'}=i{\bf e_t}\tilde{\bf W} $. Using (\ref{eq 17
2}) and (\ref{eq 6 13}) we can write solution of (\ref {eq_6_16_})
in the following form:
\begin{equation}\label{eq 6 14}\tilde{{\bf W'}}_+=-\left(1 +{\bf {e_{tr}}}{\bf
a_3}\right)k\tilde{\bf M}\exp\left\{-i{\omega_+}t+ikz\right\},
\end{equation}
and
\begin{equation}\label{eq 6 15}\tilde{{\bf W'}}_-=\left(1 -{\bf {e_{tr}}}{\bf
a_3}\right)k\tilde{\bf M}\exp\left\{-i{\omega_-}t+ikz\right\}.
\end{equation}
Note that the wave function $\tilde{{\bf W'}}_+$ describes the
positive branch of dispersion law (\ref{eq 17 2}) that
corresponds, for example, to the "antiparticle", while
$\tilde{{\bf W'}}_-$ describes the negative branch that
corresponds to the "particle" state. Besides, as it is seen the
wave functions (\ref{eq 6 14}) and (\ref{eq 6 15}) are the
eigenfunctions of spin operator \cite{b_31}:
\begin{equation}
\hat{S}_z=\frac{1}{2}~{\bf {e_{tr}}}{\bf a_3}.
\end{equation}
Indeed it is simple to check that
\begin{equation}
\hat{S}_z\tilde{{\bf W'}}=S_z\tilde{{\bf W'}},
\end{equation}
 where eigenvalue $S_z=\pm1/2$.
 It is seen that plane waves (\ref{eq 6 14}) and
(\ref{eq 6 15}) correspond to the different eigenvalues $S_z$.
Thus $\tilde{{\bf W'}}_+$
 describes "antiparticle" state with spirality $S_z=+1/2$, while $\tilde{{\bf
 W'}}_-$ describes "particle" state with spirality $S_z=-1/2$.
However in the case of massive field the plane wave
(\ref{U_res_gen}) has more complicated space-time structure.

\subsection {Nonhomogeneous equation}

Let us consider the nonhomogeneous equation corresponding to the
equation (\ref{eq_6 1a})
\begin{equation}
\label{eq_8 1}  \left(i{\bf e_{1}}\partial - {\bf
e_{2}}\vec{\nabla }-i{\bf e_{3}}m\right)\tilde{{\bf W }}=
\tilde{\bf I}.
\end{equation}
Here $\tilde{\bf I}$ is the field source. Choosing the potential
$\tilde{\bf W} $ in the form (\ref{eq_5 9}), we obtain the
following equation for the field strengths:
\begin{equation}
\label{eq_8 2} -e+i{\bf e_{1}}f-i{\bf e_{2}}g+i{\bf e_{3}}h-i\vec
E+{\bf e_{1}}\vec F+{\bf e_{2}}\vec G+{\bf e_{3}}\vec H={\bf
I}_0+\vec{\bf I}.
\end{equation}
This equation means that the strengths of this field are nonzero
only in the region of field source.

Let us consider the sedeonic source in the following form:
\begin{equation}
\label{eq_8 21}\tilde{\bf I}=-i{\bf e_{2}}4\pi\rho_L+ {\bf
e_{1}}\frac{4\pi}{c}\vec j_L .
\end{equation}
where $\rho_L$ is a volume density of charge and $\vec j_L$ is
volume density of current. In this case the equation (\ref{eq_8
2}) is rewritten as
\begin{equation}
\label{eq_8 22} -i{\bf e_{2}}g+{\bf e_{1}}\vec F=-i{\bf
e_{2}}4\pi\rho_L+ {\bf e_{1}}\frac{4\pi}{c}\vec j_L ,
\end{equation}
Applying the operator $(i{\bf e_{1}}\partial - {\bf
e_{2}}\vec{\nabla }-i{\bf e_{3}}m)$ to the equation (\ref{eq_8
22}) and separating the values with different space-time
properties we obtain the following equations for the field
strengths:
\begin{equation}
\label{eq_8 23}
\begin{array}{l}
        g=4\pi\rho_L,
\\[2mm] \displaystyle\vec F =\frac {4\pi}{c}\vec j_L,
\\[2mm] \displaystyle\partial g+(\vec\nabla\cdot\vec F)= 4\pi [\partial\rho_L+\frac {1}{c}(\vec\nabla\cdot\vec j_L)],
\\[2mm] \displaystyle[\vec\nabla\times\vec F] =\frac {4\pi}{c}[\vec\nabla\times\vec j_L],
\\[2mm] \displaystyle\partial\vec F+\vec\nabla g =4\pi[\frac {1}{c}\partial\vec j_L+\vec\nabla
\rho_L].
\end{array}
\end{equation}
Assuming charge conservation
\begin{equation}
\label{eq_8 24}\partial\rho_L+\frac {1}{c}(\vec\nabla\cdot \vec
j_L)=0
\end{equation}
we have the following gauge condition:
\begin{equation}
\label{eq_5 213} \partial g+(\vec\nabla\cdot\vec F)=0
\end{equation}
which is similar to the Lorentz gauge, but for the field
strengths.

Let us consider a stationary field generated by a scalar point
source
\begin{equation}
\label{eq_8 3} {\bf I}_0=-i{\bf{e_2}}4\pi q_L\delta(\vec{r}),
\end{equation}
where $q_{L}$  is the point charge. Then the intensity of the
scalar field is
\begin{equation}
\label{eq_8 4} g_{L}(\vec{r})=4\pi q_{L}\delta(\vec{r}).
\end{equation}
This field is non-zero only in the region of source. It indicates
that two point charges interact only if they are at the same point
of space. The interaction energy for two point charges $q_{L1}$
and $q_{L2}$ is equal

\begin{equation}
\label{eq_8 5} W_{LL}=-\frac{1}{4\pi}\int g_{L1}g_{L2}dV=-4\pi
q_{L1}q_{L2}\delta(\vec R),
\end{equation}
where $\vec R$ is the vector of distance between point charges.
Such a law of interaction is typical for leptons involved in a
weak interaction. So the $q_L$ can be interpreted as a lepton
charge.

Moreover one can suppose the interaction between $q_{B}$ and
$q_{L}$ charges due to the overlap of scalar fields $g_B$ and
$g_L$. In the case of point $q_{B}$ and $q_{L}$  the fields are
determined by the expressions (\ref{eq_5 28}) and (\ref{eq_8 4}),
so that the interaction energy is equal to:
\begin{equation}
\label{eq_8 6} W_{BL}=-\frac{1}{4\pi}\int g_{B}g_{L}dV.
\end{equation}
As a result, we get:
\begin{equation}
\label{eq_8 7}
W_{BL}=-\frac{m_{0}c}{\hbar}\frac{q_{B}q_{L}}{R}\exp
\left(-\frac{m_{0}c}{\hbar}R\right),
\end{equation}
where $R$ is the distance between  $q_{B}$ and $q_{L}$ charges.

\section{Concluding remarks}
Thus, in this paper we have presented the sixteen-component
sedeons generating associative noncommutative space-time algebra.
This algebra can be considered as the scalar-vector variant of
complexified Clifford algebra with specific commutation and
multiplication rules. The sedeonic basis elements $\bf{a}_1$,
$\bf{a}_1$, $\bf{a}_3$ are responsible for the spatial rotation,
while the elements $\bf{e}_t$, $\bf{e}_r$ and $\bf{e}_{tr}$ are
responsible for the space-time inversions.  Mathematically, these
two bases are equivalent, and the different physical properties
attributed to them are an important physical essence of our
sedeonic hypothesis.

In contrast to the Gibbs-Heaviside vector algebra the
multiplication rules for vector basis in sedeonic algebra contain
the imaginary unit (see Table 1). It enables the realization of
scalar-vector algebra whith Clifford product \cite{b_31}.
Apparently,  such possibility of vector basis multiplication was
pointed first by A. Macfarlane  \cite{b_35}. Later the similar
multiplication rules for matrix  basis were applied by W.Pauli
\cite{b_36} and P.A.M.Dirac \cite{b_37} in their spinor equations.

The important point is that the sedeonic basis elements
simultaneously play a role of the operators and space-time basis
of the wave function. From a physical point of view, this allows
us to reformulate the Klein-Gordon equation of relativistic
quantum mechanics as the wave equation for special scalar-vector
field that carries information about the kinematic properties of
quantum particles. This sedeonic Klein-Gordon equation can be
reformulated as Maxwell-like equations for the field intensities.
At the same time the sedeonic first-order Dirac wave equation can
be interpreted as the equation describing special field with zero
field intensities.

On the other hand the sedeonic Klein-Gordon equation allows
another interpretation as the wave equation for the potentials of
a force massive field. In this case the Einstein relation between
energy and momentum can be interpreted as the relation for the
energy, momentum and mass of a quantum of force field. The sources
of this field are corresponding charges $q_{B}$ and currents $\vec
j_{B}$. At the same time the sedeonic first-order wave equation
describes the special force field with zero field strengths. The
sources of this field are corresponding charges $q_{L}$ and
currents $\vec j_{L}$. We defined the concept of energy and energy
flux for the force massive field and derived an expression that
describes the energy conservation for a massive field, similar to
the Poynting theorem in electrodynamics. Based on this concept, we
have considered the interaction of point charges due to the
overlap of scalar and vector fields.

\section*{Acknowledgements}
The authors are very thankful to G.V. Mironova for kind assistance
and moral support. We also thank J. K\"{o}plinger and V. Kurin for
the useful discussions.

\end{document}